\documentclass[12pt]{article}
\usepackage{natbib,amssymb}

\begin{document}

\title{\bf Habitability properties of circumbinary planets}

\author{Ivan~I.~Shevchenko\/\thanks{E-mail:~iis@gao.spb.ru} \\
Pulkovo Observatory of the Russian Academy of Sciences, \\ 196140
Saint Petersburg, Russia}

\date{}

\maketitle

\begin{center}
Abstract
\end{center}

\noindent It is shown that several habitability conditions (in
fact, at least seven such conditions) appear to be fulfilled
automatically by circumbinary planets of main-sequence stars
(CBP-MS), whereas on Earth these conditions are fulfilled only by
chance. Therefore, it looks natural that most of the production of
replicating biopolymers in the Galaxy is concentrated on
particular classes of CBP-MS, and life on Earth is an outlier, in
this sense. In this scenario, Lathe's mechanism for the tidal
``chain reaction'' abiogenesis on Earth is favored as generic
for CBP-MS, due to photo-tidal synchronization inherent to them.
Problems with this scenario are discussed in detail.

\bigskip


\bigskip

\section{Introduction}

Certain conditions should be satisfied for life, as present on the
Earth, to emerge and be sustained on a planet; such conditions
concern the insolation level~\citep{H60}, seasons~\citep{OB16},
climate stability~\citep{LJR93}, tidal phenomena~\citep{L04},
protection from XUV radiation and stellar wind~\citep{M13}, active
tectonics~\citep{VBG14}, and presence of water~\citep{OML90},
among others. The importance of each of these factors can be
quantified only approximately, but it is at least known that their
intensity and variation as present on the Earth are suitable for
life.

As discussed below, the apparent fulfillment of all the stated
conditions on the Earth is due to an overlap of lucky chances. One
may ask: are there planets that exist where these conditions are
satisfied automatically (generically)? In a more technical
formulation, are there generic biochemical reactors that exist to
produce replicating biopolymers, as, e.g., analogously, stars can
be considered as reactors for producing metals?

Recently, \cite{M13,M15,M15b} and \cite{ZMC16}, analyzing
habitability conditions in circumbinary planetary systems, came to
the conclusion that planets orbiting stellar binaries with
particular parameters might be exceptionally good, from an
astrophysical viewpoint (concerning the protection of life from
XUV radiation and stellar wind), as life habitats: mutual tides in
the stellar main-sequence binaries with orbital periods greater
than $\sim$10~d radically suppress the magnetic dynamo mechanism,
thus reducing the chromospheric activity and the life-hostile
extreme UV radiation and stellar wind. This suppression is
favorable for life to emerge and survive on the planets orbiting
such stars. Therefore, as follows from these astrophysical
arguments, ``life may even thrive on some circumbinary
planets''~\citep[p.~391]{M15}.

Here we enforce this conclusion, arguing that there are, in fact,
a number of major life-supporting conditions that arise naturally
on many circumbinary planets of main-sequence stellar binaries
(abbreviated ``CBP-MS'' in what follows).

On the other hand, we argue and emphasize that the fulfillment of
these conditions on the Earth is purely accidental. Therefore, the
CBP-MS are possibly the main ``life-breeders'' in the Galaxy,
while life on Earth is an accidental exception, as the Earth
accidentally mimics a typical CBP-MS in several life-favorable
respects.

\section{Insolation}
\label{sec_Insolation}

The observational data on the recently discovered CBP-MS ({\it
Kepler}-16b, 34b, 35b, and others) testify that most of them move
in the orbits closely encircling central zones of orbital
instability around their host binaries \citep{D11,W12,W14,PS13}.
According to Popova \& Shevchenko (2016a, Table 3), the observed
planets concentrate in the resonance cells between orbital
resonances 5/1 and 6/1 or 6/1 and 7/1 (designated as resonance
cells 5/1--6/1 and 6/1--7/1 in what follows) with the binary.
Their location is, therefore, quite predictable.

The presence of the central zone of orbital instability (chaotic
zone) is automatic for any binary with a large enough mass ratio
of companions: in fact, the gravitating binaries with components
of comparable masses possess circumbinary zones of dynamical
chaos, as massive numerical simulations
\citep{D84,D86,RD88,DFF89,HW99} and theoretical celestial
mechanics \citep{S15} in the framework of the restricted
three-body problem tell us.

The extent of the chaotic zone around a system of two
gravitationally bound bodies was estimated analytically in
\cite{S15}, based on Chirikov's resonance overlap
criterion~\citep{C79}. The binary's mass ratio, above which such a
chaotic zone is universally present, was also estimated. These
analytical results are in agreement with the modern data on the
orbits of CBP-MS~\citep{S15,PS16b}.

Central cavities of analogous origin are observed in
protoplanetary disks, which contain planetesimals, dust, and gas;
the gas is present in the initial stages of the disk evolution. In
contrast to disks of single stars, circumbinary disks have large
central cavities. The existence and possible characteristics of
such cavities in gaseous circumbinary disks were first considered
analytically by \cite{AL94,AL96a}. The central cavity exists
irrespective of the gas content in the disk.

All observed host stars of CBP-MS belong to a particular part of
the overall period distribution of stellar binaries: their period
range is 7--40~d~\citep{MMF15}. The ``dearth'' of planets around
the shorter-period main-sequence binaries is naturally explained
as a consequence of the Lidov--Kozai
effect~\citep{MMF15,HPP16,ML15}: if a third distant stellar
companion is present, the Lidov--Kozai mechanism in concert with
tidal friction shrinks the inner binary, and this shrinking
perturbs the planets. They either escape or fall on the stars, or
even their formation is prevented. In fact, the Lidov--Kozai
mechanism in concert with tidal friction is considered to be
responsible for the formation of most stellar binaries with
periods below 7~d~\citep{MMF15,WF15}. Thus, the planet-hosting
stellar binaries with periods less than $\approx$7~d are lacking,
naturally due to the Lidov--Kozai mechanism in stellar triples.

On the other hand, when CBP-MS are formed, they migrate toward the
central stellar binary and stall at the outer border of the
chaotic zone around the binary because there is no more matter to
cause the migration~\citep{PN07,M12,PLT12}. (Formation and
migration issues are considered in more detail in
Subsection~\ref{subsec_Formation}.)

If the period of the central binary is $\sim$10~d, then the
minimum period of the stalled planet is $\sim$50~d because the
resonance cell 5/1--6/1 with the binary is typically the closest
(to the binary) stable one: it is located at the border of the
central chaotic region around the binary; see \cite{S15}.
Therefore, for the Solar-like star binaries with periods
$\sim$10--100~d the border-line (``leading'') CBP, or the planets
trapped in outer low-order mean motion resonances with this
leading CBP, are automatically placed close to the habitable zone
(HZ) or inside it.

This is made as an order of magnitude estimate, using a rough
analogy with the location of the Earth in the Solar system. Of
course, habitable zone science, in its progress, now allows one to
make much more rigorous estimates~\citep{C14,KH13,KRK13,MH14}. In
what follows, we use equations from \cite{KH13}, in particular.

Most of the observed CBP-MS are indeed close to the habitable zone
\citep{W14}. As we have just seen, this is not just a mere
coincidence, but an inevitable consequence of generic dynamical
and physical effects. Concerning Earth, its life-favorable surface
temperature is, in fact, fortuitous: in contrast to the
circumbinary systems, there is no such a straightforward mechanism
in the Solar system known to stall a planet at an appropriate
orbital distance from the Sun; the presence of the Earth (and
marginally Mars) in the Solar habitable zone is accidental, in
this sense. However, one should admit that the absence of a known
mechanism does not necessarily mean that it is actually absent. In
numerical simulations, there do exist planetary formation
scenarios where the overall architecture of the Solar system, with
rocky planets residing in the habitable zone, is reproduced; see
\cite{BMI16}. We discuss formation issues further on in
Subsection~\ref{subsec_Formation}.

The main-sequence binary stars in the Galaxy have a rather wide
distribution of periods \citep{DM91}, with a median value of
$\approx$180~yr. Therefore, if one considers the overall
population of the main-sequence binary stars, then the suitable
lower bound of insolation is not provided automatically for the
habitability of CBP-MS. However, there seems to exist a physically
distinct stellar subpopulation that does provide such a bound.
This is the population of so-called ``twin binaries'': the
near-equal mass binaries (namely the binaries with mass ratios
from $\sim$0.8 to 1) forming a statistical excess at short orbital
periods~\citep{HMU03,Lu06,SO09}. For the twins, the median period
is $\sim$7~d, and the upper cutoff of the period distribution is
at $\approx$43~d~\citep{Lu06,SO09}. Therefore, at the cutoff, the
CBP orbiting at the border of the central chaotic zone around the
binary (in resonance cell 5/1--6/1 or 6/1--7/1) would have the
orbital periods $\sim$200--300~d, quite close to the inner border
of the habitable zone of a double Solar-like star.

Twins seem to be physically distinct, in their formation process,
from other binaries~\citep{HMU03}: the binary components could
form in situ (and the process could be followed by accretion from
a common gaseous envelope, equalizing the masses of the
components), whereas the binaries with smaller mass ratios may
acquire shorter periods by migration in the circumbinary disk.

Note that, in twin population statistics, the mass ratio of
stellar companions may have a range as wide as
$\approx$0.8--1.0~\citep{HMU03}; therefore, the ratios of stellar
radii can be as low as $\approx$0.8, close to the ratio allowing
for the ``optimal eclipse'' (see Section~\ref{sec_Eclipses}).

Quantitatively, what is the percentage of binaries that may give
birth to habitable CBP-MS? Nowadays, only very rough estimates can
be made. According to the distribution of periods $P$ of nearby
G-dwarf binaries given in Duquennoy \& Mayor (1991, Figure~7), 13
of 82 binaries in the sample ($\approx$16\%) are short-period ($P
< 100$~d). On the other hand, according to the same database, a
much larger percentage is observed, e.g., in the Hyades field,
where short-period binaries constitute $\approx$44\% of the total
sample. Thus, it might be no exaggeration to say, that, by the
order of magnitude, $\sim$10\% of all main-sequence binaries are
suitable, at least among G-dwarfs.

Let us estimate the sizes of habitable zones for twin binaries.
According to \cite{UJS03} and Kane \& Hinkel (2013, Equations
(2)--(5)), the stellar fluxes at the inner and outer borders of
the habitable zone of a single star are given by

\begin{equation}
S_\mathrm{inner} = 4.190 \cdot 10^{-8} T_\mathrm{eff}^2 - 2.139
\cdot 10^{-4} T_\mathrm{eff} + 1.268 , \label{S_inner}
\end{equation}

\begin{equation}
S_\mathrm{outer} = 6.190 \cdot 10^{-9} T_\mathrm{eff}^2 - 1.319
\cdot 10^{-5} T_\mathrm{eff} + 0.2341 , \label{S_outer}
\end{equation}

\noindent and the locations of the inner and outer borders of the
habitable zone are given by

\begin{equation}
r_\mathrm{inner} = (L/S_\mathrm{inner})^{1/2} , \qquad
r_\mathrm{outer} = (L/S_\mathrm{outer})^{1/2} , \label{r_HZ}
\end{equation}

\noindent where the radii $r_\mathrm{inner}$ and
$r_\mathrm{outer}$ are in astronomical units, stellar luminosity
$L$ in Solar units, and effective temperature $T_\mathrm{eff}$ in
Kelvins.

For a twin binary consisting of companions with luminosity
$L_\mathrm{*}$ and effective temperature $T_\mathrm{eff*}$, the
locations of the inner and outer borders of the circumbinary
habitable zone can be roughly estimated using the hierarchical
approximation, by setting $L=2L_\mathrm{*}$ and
$T_\mathrm{eff}=T_\mathrm{eff*}$ in the given equations. Using
data from Kaltenegger \& Traub (2009, Table~1), one finds that the
circumbinary habitable zone overlaps completely with resonance
cells 5/1--6/1 and 6/1--7/1 of a central binary with period 7~d
(the median period for twins) if the twin is of M2V type (doubled
luminosity $L=2 \cdot 0.023=0.046$, doubled mass $M=2 \cdot
0.44=0.88$ in Solar units, $T_\mathrm{eff}=3400$~K), or M3V type
($L=2 \cdot 0.015=0.030$, $M=2 \cdot 0.36=0.72$,
$T_\mathrm{eff}=3250$~K), or M4V type ($L=2 \cdot 0.0055=0.011$,
$M=2 \cdot 0.20=0.40$, $T_\mathrm{eff}=3100$~K). For other
spectral classes, there is not even any partial overlap.

For the twins with the cutoff period $\sim$40~d, the outer edge of
the resonance cell 6/1--7/1 is close to the inner edge of the
habitable zone even for Solar-like yellow dwarf twins
($L=2L_\mathrm{Sun}=2$, $M=2M_\mathrm{Sun}=2$,
$T_\mathrm{eff}=T_\mathrm{eff(Sun)}=5770$~K).

Due to the general correlation of the planetary and host-star
masses \citep{RSM07,R08}, CBP in M-dwarf systems should be
generally much smaller than Neptune-like CBP in the {\it Kepler}
systems; in fact, they may be Earth-like as, e.g., the Earth-like
planets observed in the Trappist-1 single-star system
\citep{GTD17}. The problem of planetary masses is discussed in
more detail in Subsection~\ref{subsec_Formation}.

The observed host stars of {\it Kepler} CBP are mostly Solar-like
stars, not M-dwarfs. Therefore, let us consider the case of
Solar-like star twin binaries in more detail. Today, we are aware
of the existence of several {\it Kepler} systems that resemble
such twins ({\it Kepler}-34, 1647), or have at least one component
that is Solar-like ({\it Kepler}-38, 47, 453). All known CBP are
Neptune-like or Jupiter-like giant planets, and most of them
reside in orbits close to the central chaotic zone, as mentioned
above. The fact that only giant CBP have been discovered up to now
seems to be a natural observational bias, and the existence of
less massive planets in orbits outside of the observed ``leading''
CBP is in no way precluded. (Inner planets are precluded as they
would reside in the chaotic zone.) What is more, the existence of
such objects looks plausible. Indeed, in the Solar system, the
outer giant planet, Neptune, shepherds Pluto and plutinos in 3/2
orbital resonance with Neptune, and a number of other
trans-Neptunian objects (TNO) in higher-order half-integer
resonances. According to \cite{G12}, the population of TNO in 5/2
resonance with Neptune is estimated to be as large as that in 3/2
resonance.

As for the outer (presumably terrestrial) CBP, they can be
expected to reside in three-body resonances with the central
binary and the leading giant planet, forming stable
configurations, as discussed below. (On the three-body resonances
in the Solar system, see \citealt{SS13}). Besides, a plausible
orbital architecture of CBP can be expected to resemble that
observed in the satellite system of the Pluto--Charon binary,
where Styx, Nix, Kerberos, and Hydra all move in the orbits close
to integer resonances with the Pluto--Charon binary
\citep{BGT13,BSJ15}, and the system is dynamically closely-packed.

The inference that various CBP in the habitable zone outside the
leading CBP may indeed actually exist follows from the example of
the {\it Kepler}-47 system, where at least two CBP {\it
Kepler}-47b and {\it Kepler}-47c are observed
\citep{OWC12a,KMH13}, and there is enough dynamical space between
their orbits to harbor more planets \citep{KS14,HHK15}. These
latter planets should be much smaller than b and c (because they
are not observed at present) and, thus, can be super-Earths or
even Earth-like.

Among the binaries hosting {\it Kepler} CBP, according to Welsh et
al.\ (2012, Table~1), there are two twin binaries, namely {\it
Kepler}-34 and {\it Kepler}-35. Let us consider first {\it
Kepler}-34 as a Solar-like star twin binary prototype. Using
formulas~(\ref{r_HZ}) and observational orbital and physical data
from Welsh et al.\ (2012, Table~1) and \cite{HK13}, one can
readily find that the inner and outer radii for this binary's
habitable zone are equal to $\approx$1.4 and $\approx$2.7~AU,
corresponding to orbital periods of 410 and 1150~d. These values
are respectively $\approx$1.4 and $\approx$4.0 times greater than
the period of the planet {\it Kepler}-34b. Therefore, the outer
orbital resonances 3/2, 2/1, 5/2, 3/1, and 7/2 with the observed
giant planet are all within the HZ. These are the resonances where
habitable rocky planets (Earth-like ``plutinos,'' ``twotinos,''
etc.), shepherded by the leading CBP {\it Kepler}-35b, may reside.
An analogous calculation for {\it Kepler}-35 shows that its HZ
ranges approximately from resonance 2/1 to resonance 6/1 with the
leading planet {\it Kepler}-35b, i.e., the HZ starts with
``twotinos.''

Among these outer resonances, 2/1 is of particular interest.
Indeed, the observed {\it Kepler} CBP tend to concentrate in the
centers of resonance cells bounded by integer resonances with the
central binary \citep[figure~4]{S15}. The centers of the cells
correspond to the half-integer resonances $m/2$ with the central
binary; therefore, a planet in 2/1 resonance with an observed
leading CBP would tend to be additionally in the integer resonance
$m/1$ with the central binary. Thus, the whole system would be
close to a three-body resonance. This fact is interesting from the
viewpoint of the long-term stability of such a configuration; the
stability deserves a further study. (On the importance of
three-body resonance in closely-packed planetary systems, see
\citealt{Q11}.)

One should outline that it is not only in {\it Kepler}-34 and {\it
Kepler}-35, but also in the overall population of Solar-like star
twin binaries that the twotinos of the leading CBP (``twotino
tatooines'') are more or less automatically placed in the
insolation HZ. This is due to the fact that such stellar binaries
have a suitable median orbital period, as discussed above.

An important observational circumstance is that among the seven
first discovered {\it Kepler} CBP, the planets not belonging to
the stellar HZ or its extended versions are all {\it inside} the
inner edge of the HZ \citep[Table~1]{W14}; therefore, all of these
giant planets may shepherd smaller outer planets in the HZ.

In conclusion, at least two habitable niches, where the habitable
level of insolation is provided more or less automatically, can
exist: (1)~the leading CBP of M-dwarf twin binaries and (2)~the
outer terrestrial bodies shepherded by the leading CBP of
Solar-like star twin binaries.

At present, the prospects for an observational discovery of a
terrestrial planet belonging to any of these two classes are
rather vague because its transit signal is by two or three orders
of magnitude less than that of the discovered leading CBP;
besides, in the second class, its influence on the transit timing
variations of the leading CBP is also small.

Relevant issues concerning the formation and migration of
terrestrial planets of single and binary stars are considered
further on in Subsection~\ref{subsec_Formation}.

\section{Seasonal variations}
\label{sec_Seasons}

\cite{OB16} showed that the inherent long-term instability of
planetary climates can be quenched if seasonal variations are
present. In other words, seasonal variations might be a necessary
condition for a habitable climate to be maintained.

The seasonal analogy between the Earth and a typical CBP-MS first
of all means that the amplitudes of surface temperature variations
are similar. For the CBP-MS, the life-favorable amplitude
($\sim$10\%, as on the Earth) of the temperature variation arises
generically, as the following simple calculation for a twin binary
shows, even if the planet has zero tilt of the rotation axis.

The relative amplitudes of the surface temperature
$T_\mathrm{max}/T_\mathrm{min}$ and stellar radiation flux
$F_\mathrm{max}/F_\mathrm{min}$, related by the equation

\begin{equation}
T_\mathrm{max}/T_\mathrm{min} \sim
(F_\mathrm{max}/F_\mathrm{min})^{1/4} , \label{SB_law}
\end{equation}

\noindent are derived from the dependence of the flux $F$ on the
phase angle $\varphi$, given by the formula

\begin{equation}
F(x) \propto r_1^{-2} \cos(a r_1^{-1} \sin \varphi ) + r_2^{-2}
\cos(a r_2^{-1} \sin \varphi ) , \label{inso}
\end{equation}

\noindent where $r_1^2 = a^2+b^2 - 2 a b \cos \varphi$, $r_2^2 =
a^2+b^2 + 2 a b \cos \varphi$, $a$ is the semimajor axis of the
central twin, $b$ is the barycentric radius of the planetary
orbit, and $\varphi$ is the angle between the planet's barycentric
radius-vector and the ``star~1 -- star~2'' direction. Of course,
using the Stefan--Boltzmann law in Equation~(\ref{SB_law}) is a
kind of oversimplification, but here we make only rough estimates.

Let us see what is the relative amplitude of the surface
temperature variations at the orbital radius corresponding to
period ratio $\sim$6/1 (6/1 resonance between resonance cells
5/1--6/1 and 6/1--7/1). Setting $a=1$, $b=6^{2/3} \approx 3.302$,
from Equation~(\ref{inso}) one has $\approx$1.50 for the ratio of
the maximum and minimum fluxes, and $\approx$11\% for the relative
amplitude of the corresponding temperature variations. For period
ratios 5/1 and 7/1, the temperature amplitude is the same, with
accuracy better than 2\%.

Concerning Earth, the favorable range of its surface temperature
variations is conditioned by an accidental, quite high value of
the obliquity (most probably caused by a giant impact; see
\citealt{W93}) of Earth's equator to the ecliptic plane. Note that
the generic tilt of a planet in a relatively low orbit is equal to
zero, as in the case of Mercury and Venus, because this is a
natural outcome of the tidal spin--orbit evolution of orbiting
bodies (see, e.g., \citealt{WPM84}).

However, there exists a difference in the timescale: the seasonal
variations on CBP-MS generically occur on a month scale (in
contrast to the Earth); however, if a CBP-MS is tilted, the annual
harmonic is also present.

\cite{OB16} designate three solutions to the heat balance
equation: the ``snowball-planet point'' (low effective
temperature), the ``habitable-planet point'' (suitable effective
temperature), and the ``desert-planet point'' (high effective
temperature). The habitable point is inherently unstable. However,
it can be stabilized by high-frequency periodic driving, provided
(on the Earth) by seasonal variations. It is stabilized in the
same way as the inverted rigid pendulum is stabilized by
high-frequency periodic oscillations of the suspension point
\citep{Ste08,Kap51,Kap54}. For the pendulum, the phenomenon was
demonstrated in real physical experiments; see \cite{Kap51,Kap54}.

The exact value of the perturbation frequency is unimportant, as
long as the frequency is great with respect to the long-term
climate oscillation (Ice Ages) frequency, and the perturbation
amplitude remains the same by an order of magnitude. Therefore,
whether the perturbation is ``obliquity based'' (with the period
of several months) or ``binary based'' (with the period of several
weeks), it has the same stabilizing effect, as long as the
amplitudes are the same. However, if both ``obliquity-based'' and
``binary-based'' oscillations of comparable amplitudes are
present, the combined effect is not obvious and should be explored
in numerical experiments.

\section{Climate stability}
\label{sec_Climates}

\cite{LJR93} and \cite{NL97} argued that the current obliquity of
the Earth is secularly stable due to the presence of the Moon: the
Moon-caused precession of Earth's spin axis is rapid enough to
prevent the chaotic diffusion between relevant spin--orbit
resonances, as they are more widely separated in the phase space,
due to this precession.

If the Moon were absent, the Earth would suffer large variations
of its obliquity (between $0^\circ$ and $85^\circ$), entailing
catastrophic variations of climate \citep{LJR93,NL97}. Conversely,
the obliquity of the actually Moonless Mars varies in the range
0$^\circ$--60$^\circ$ \citep{LR93}, and this is at least one of
the causes of its sterility, although it is marginally inside the
Solar habitable zone. With the Moon, Earth's obliquity stays
within the range from $22.1^\circ$ to $24.5^\circ$~\citep{LBC12}.

The issue, however, remains highly disputable. \cite{LBC12} and
\cite{LB14a} argue on the basis of results of massive numerical
experiments that the chaotic diffusion rate in the obliquity of a
Moonless Earth is low enough for the development of life to be
successful, and the long-term habitability is not precluded.

Nevertheless, one may speculate that the presence of a second star
may be important as a stabilizing factor for the obliquity of the
planetary rotation axis (vital for the climate stability), similar
to the role played by the Moon. The binarity of a host star as the
attitude stabilizer has not yet been explored. However, one may at
least say that it does not induce the attitude instability. This
can be argued as follows. The planar rotations-oscillations of an
asymmetric satellite in a fixed elliptic orbit around a single
primary body are described by the Beletsky equation~\citep{B63}. A
generalized (circumbinary) version of the Beletsky equation,
taking into account the torque exerted on a rotating body by an
inner massive satellite of the primary body, has been derived by
\cite{C15}, in connection with the problem of rotation of minor
outer satellites in the Pluto--Charon system. Adapting the results
of \cite{C15}, one sees that the major spin--orbit resonances
overlap and the rotational chaos becomes global if

\begin{equation}
\left( 3 \frac{B-A}{C} \right)^{1/2} \gtrsim \frac{1}{2} \left(
\frac{n_\mathrm{b}}{n_\mathrm{p}} - 1 \right) , \label{ch_rot}
\end{equation}

\noindent where $A < B < C$ are the principal moments of inertia
of the planet, and $n_\mathrm{b}$ and $n_\mathrm{p}$ are the mean
motions (orbital frequencies) of the stellar binary and the
planet, respectively; the orbits of the binary and the planet are
assumed to be circular and the masses of the binary components
comparable to each other. If condition~(\ref{ch_rot}) does not
hold, the chaotic layers are thin and the motion is mostly (with
respect to the initial conditions) regular.

CBP of twin binaries are orbitally stable if
$n_\mathrm{b}/n_\mathrm{p} \gtrsim 4$--5; see \cite{HW99} and
\cite{S15}. Therefore, according to criterion~(\ref{ch_rot}), for
such objects, there is no macroscopic rotational chaos in the
planar setting of the problem, even if $(B-A)/C$ is as great as
$\sim$0.1. For rocky Earth-like planets, the quantity $(B-A)/C$ is
expected to be not greater than $\sim$$10^{-5}$; see \cite{C15}
and references therein. Therefore, the circumbinary rocky planets
are unlikely to rotate chaotically in this planar setting of the
problem. The instability with respect to tilting the spin axis
deserves a further study.

According to a recent analysis by \cite{C16}, when the precession
rate of a CBP's spin is much less than the precession rate of its
orbit (the situation typical for CBP, as argued by \citealt{C16}),
any significant oscillations in its obliquity are not expected, at
least in the framework of the full three-body problem.

Apart from the planetary obliquity variations, climates may suffer
Milankovitch cycles. Frequent Ice Ages (on the timescale of order
1000~yr) were predicted by \cite{F16} for the CBP with typical
orbital parameters as observed for {\it Kepler} systems. The
climates of such planets were shown to suffer short-timescale
Milankovitch cycles, which might trigger Ice Ages. Frequent Ice
Ages do not seem to be favorable for long-term habitability, but
this issue deserves a further study. This is a complex problem,
and Milankovitch cycles are not necessarily coupled to Ice Ages;
see \cite{G89}, \cite{McD90}, and \cite{S02}.

\section{Tides}
\label{sec_Tides}

The importance of tides for abiogenesis is at least twofold:
first, they produce periodic wetting and drying of beaches, and
this process is favorable for life in several respects
\citep{L04}; second, they provoke plate tectonics, also favorable
for life in several respects, in particular by allowing important
chemicals produced deeply in planetary interiors to come to its
surface \citep{VBG14}. According to \cite{L04}, life on Earth, in
fact, has a tidal origin.

As discussed in Section~\ref{sec_Insolation}, most of the observed
CBP-MS reside in resonance cells delineated by the chaotic bands
corresponding to 5/1, 6/1, and 7/1 mean motion resonances with the
central binary. Therefore, the typical ratio of the orbital
periods of the planet and the binary is $\sim$6. Since the stellar
binary components usually have comparable masses, the Fourier
expansion of the time-varying gravitational potential acting on
the planet possesses a dominating term with the doubled frequency.
If the masses are equal, then the period of neap/spring tides on
the planet effectively doubles, i.e., the ratio of the tide
frequency to the planet orbital frequency is $\sim$12, analogous
to what we have (at present) on the Earth subject to the Lunar
tides.

The repeatedly dried and wetted tidal pools are thought to be a
possible place of origin of self-replicating biopolymers, as such
pools provide favorable conditions for concentrating organic
molecules (see \citealt{L04} and references therein).
\cite{L04,L05,L06} proposed a theory of abiogenesis, based on the
tidal ``boosting'' of the biomolecule production in near-shore
lakes and ponds. The essence of the theory consists of the
hypothesis that periodic concentration/dissociation of complex
molecules leads to a ``chain reaction'' in the production of
specific nucleic acids.

An open question with this theory is that it postulates frequent
and high tides, not characteristic of the modern Earth and,
presumably, for the early Earth as well \citep{L06,VRD06}.
Besides, the rates of concentration/dissociation are not
determined quantitatively. The rates of evaporation in modern
lakes are of the order $\sim$1~cm/day maximum \citep{J10}. They
are controlled by the processes of Solar energy storage in the
lakes' water. Therefore, much more time than just several hours,
as in Lathe's original theory~\citep{L04,L05,L06} might be needed
to concentrate the ``bio-soup.''

A high Solar flux sufficient for the rapid-enough evaporation may
extinct the emerging life altogether. Therefore, some kind of
photo-tidal synchronization is needed. Such a synchronization is
automatically maintained on CBP-MS, where ``seasonal'' variations
of insolation are synchronized with neap/spring tides.

It is well known that the Solar and Lunar tides raised on the
Earth have rather similar amplitudes. The ratio of the amplitudes
of the Solar and Lunar tides is equal to

\begin{equation}
\frac{m_\mathrm{Sun}}{m_\mathrm{Moon}}
\left(\frac{a_\mathrm{Moon}}{a_\mathrm{Sun}}\right)^3 \approx 0.46
, \label{eq_amplitudes}
\end{equation}

\noindent as given by formula~(4.17) in \cite{MD99}.

It follows directly from Equation~(\ref{eq_amplitudes}) that the
relative amplitudes of the Solar and Lunar tides on the Earth are
roughly the same as the relative amplitudes of the tides raised on
a typical CBP-MS by its host stellar binary's components. Indeed,
in the latter case one has

\begin{equation}
\frac{m_1}{m_2} \left(\frac{a_2}{a_1}\right)^3 \approx
\frac{m_1}{m_2} \sim 1 , \label{eq_a_cb}
\end{equation}

\noindent because the planet's radial distances from the binary
components $a_1 \approx a_2$, and the masses of the main-sequence
components $m_1 \sim m_2$.

The absolute amplitudes of the Lunar and Solar tides are equal to
$0.36$~m and $0.16$~m, respectively~\citep{MD99}. For a CBP in the
habitable zone of a stellar binary whose most massive component is
Solar-like, the absolute and relative amplitudes of the tides
raised by the binary components are of the same order as on the
Earth if the mass ratio $m_1/m_2$ of the companions is $\sim$2.
Tantalizingly, the condition for ``optimal eclipses'' (see
Section~\ref{sec_Eclipses}) also requires inequality of the
stellar masses by a factor of $\sim$1.5--2.

In the polymerase chain reaction (PCR), periodic cycling between
low and high temperatures (50--100$^\circ$C) drives the
exponential amplification of the concentration of DNA molecules:
at low temperature, association is promoted and complementary
strands are synthesized (the molecule number doubles), whereas at
high temperature, the duplex strands dissociate. When the
temperature is low again, the molecule number again doubles, and
so on (for details, see \citealt{L04,L05}). In Lathe's tidal chain
reaction (TCR), the water salinity, not temperature, is
periodically driven, leading to the same exponential amplification
of nucleic acids \citep{L04,L05}. At low salinity, association is
promoted and complementary strands are synthesized, and, at high
salinity, the duplex strands dissociate.

In the case of the Earth--Moon system solely the salinity factor
operates because one of the tide-raising bodies does not radiate
practically. The case of CBP-MS is different: as a tide-raising
body (any component of the stellar binary) approaches the planet,
both the tide driven by this star and radiation flux from this
star rise, and therefore, salinity decreases and temperature
increases in concert, and vice versa. Thus, PCR and TCR conditions
are simultaneously satisfied, and the whole process is expected to
be more effective. However, a problem is that the reaction product
is unstable and decays quite rapidly, as discussed in detail in
\cite{L05}. Perhaps encapsulation in protocells increases
survivability; see Subsection~\ref{subsec_aaft}. Another problem
is that the photo-tidal synchronization may result in the UV
radiation sterilizing tidal pools during a drying phase. This is a
complex problem, depending on the spectra of the host stars and
the light-absorption properties of atmospheres. What is more, the
periodically variable UV radiation may be, on the contrary,
favorable for abiogenetic chemical reactions, as discussed in
Subsection~\ref{subsec_aaft}.

Both the Earth's tilt and Moon's presence are due to chance, as
follows from modern cosmogonical theories~\citep{L97,LBC12}.
Conversely, on the CBP-MS the synchronism of insolation and tidal
variations (the photo-tidal synchronism) with suitable periods is
maintained automatically, as shown above, and, what is more, the
amplitudes of these variations can be basically the same as on the
Earth.

In conclusion, Lathe's tidal mechanism \citep{L04,L05} of
abiogenesis, if active in reality, seems to be generic for CBP-MS
and solely accidental for Earth; then, life on Earth is favored by
the complete (both in period and amplitude) similarity of the
neap/spring tidal conditions on CBP-MS and the Earth.

\section{Active tectonics}
\label{sec_Active_tect}

Active tectonics facilitate the habitability of Earth-like
planets: ``terrestrial life depends on heat-driven plate tectonics
to maintain the carbon cycle and to moderate the greenhouse
effect''~\citep[p.~1888]{VBG14}. On the primitive Earth,
underwater volcanic processes provided energy and material for the
synthesis of organic compounds; due to large pressure and
temperature gradients, underwater volcanoes or hydrothermal
systems favor the survival, in their close vicinities, of the
organic compounds formed~\citep{M76,K12}.

The compositional diversity of terrestrial exoplanets, according
to numerical models, can be rather broad; see \cite{CBO12} and
references therein. Taking into account the effects of giant
planet migration leads to the conclusion that planets with
Earth-like composition can be ubiquitous and, what is more, the
delivery of water to terrestrial planets can be favored and
enhanced \citep{CBO12}. Therefore, the tidal dissipation parameter
$Q$ and Love number $k_2$ for terrestrial exoplanets can be
expected, in many cases, to be mostly the same as for the rocky
planets in the Solar system.

The habitability may require a heat source driven by tidal
friction: without such a mechanism, a rocky planet normally cools
down on the timescale of 10~Gyr \citep{VBG14}. The presence of a
perturber, producing tides, may be required for the heating. For
the tidal friction to be active, the perturber should maintain the
planet's orbital eccentricity. Tantalizingly, the orbital
eccentricities of CBP are periodically forced on secular
timescales, as shown by \cite{MN04}.

The central binary induces forced eccentricity in circumbinary
orbits; the eccentricity is forced periodically on secular
timescales \citep{MN04}. Thus, the orbits cannot sustain permanent
circularity. In the case of zero eccentricity of the central
binary, the forced eccentricity is given by the formula

\begin{equation}
e_\mathrm{p} = \frac{3}{4} \frac{m_1 m_2}{(m_1 + m_2)^2}
\left(\frac{a_\mathrm{b}}{a_\mathrm{p}}\right)^2 = \frac{3}{4}
\frac{m_1 m_2}{(m_1 + m_2)^2}
\left(\frac{T_\mathrm{b}}{T_\mathrm{p}}\right)^{4/3}
\label{eq_e_b}
\end{equation}

\noindent \citep{PLT12}, where $m_1 \geq m_2$ are the masses of
the binary components; $a_\mathrm{b} \ll a_\mathrm{p}$ are the
semimajor axes of the binary and the particle, respectively; and
$T_\mathrm{b}$ and $T_\mathrm{p}$ are their orbital periods. From
formula~(\ref{eq_e_b}), for a stellar twin binary ($m_1 = m_2$)
one has

\begin{equation}
e_\mathrm{p} = \frac{3}{16}
\left(\frac{a_\mathrm{b}}{a_\mathrm{p}}\right)^2 = 0.1875
\left(\frac{T_\mathrm{b}}{T_\mathrm{p}}\right)^{4/3} .
\label{eq_e_b_tw}
\end{equation}

\noindent For a circumbinary object in the orbital range between
resonances 5/1 and 7/1 with the central binary, this formula gives
$e_\mathrm{b} = 0.022$--$0.014$, similar to the current
eccentricity of Earth's orbit ($\approx 0.0167$). Again, Earth's
value is accidental in the sense that it is not an outcome of any
known automatic process, contrary to the CBP's case. The
eccentricity similarity  means that the heating conditioned by the
eccentricity is also similar in both cases, i.e., it is much less
than the radiogenic heating on the Earth (where the current heat
flow is 0.065 and 0.1~W~m$^{-2}$ through the continents and the
ocean crust, respectively; \citealt{ZAC07}).

The tidal heating flux through the planetary surface is estimated
as

\begin{equation}
h = \frac{63}{16 \pi} {\cal G}^{3/2} M_\mathrm{s}^{5/2}
R_\mathrm{p}^3 Q_\mathrm{p}'^{-1} a_\mathrm{p}^{-15/2}
e_\mathrm{p}^2 \label{eq_h_tidal}
\end{equation}

\noindent \citep{JGB08,BJG09}, where ${\cal G}$ is the
gravitational constant, $M_\mathrm{s}$ is the stellar mass,
$a_\mathrm{p}$ and $e_\mathrm{p}$ are the semimajor axis and
eccentricity of the planetary orbit, $R_\mathrm{p}$ is the
planetary radius, and $Q_\mathrm{p}' = 3Q_\mathrm{p} /(2
k_\mathrm{p})$, where $Q_\mathrm{p}$ and $k_\mathrm{p}$ are the
planetary tidal dissipation parameter and Love number,
respectively. In our hierarchical model, $M_\mathrm{s}=m_1+m_2$.
Combining Equation~(\ref{eq_h_tidal}) with
Equation~(\ref{eq_e_b_tw}), one has

\begin{equation}
h = \frac{567}{128} \pi^4 R_\mathrm{p}^3 {\cal G}^{-1}
Q_\mathrm{p}'^{-1} f^{-23/3} T_\mathrm{b}^{-5} , \label{eq_h_tres}
\end{equation}

\noindent where $f$ ($\equiv T_\mathrm{b}/T_\mathrm{p}$) is the
ratio of the orbital periods of the planet and the binary. Note
that, quite unexpectedly, the combining of
Equation~(\ref{eq_h_tidal}) with Equation~(\ref{eq_e_b_tw}) has
cancelled the stellar mass; therefore, formula~(\ref{eq_h_tres})
is valid for any type of the central twin binary, be it composed,
say, of M-dwarfs or Solar-like stars.

According to heuristic estimates in \cite{BJG09}, the heating
rates less than 0.04~W~m$^{-2}$ and greater than 2~W~m$^{-2}$
imply non-habitability. Taking $T_\mathrm{b}=7$~d (the median
period of stellar twin binaries; see
Section~\ref{sec_Insolation}), $R_\mathrm{p}=2 R_\mathrm{Earth}$
as for a super-Earth, and $Q_\mathrm{p}' = 500$ as justified in
\cite{BJG09}, one finds that the tidal heating for such a planet
provides habitability at $f= 1.8$--$3.2$, i.e., approximately in
the range of resonances from $2/1$ to $3/1$ with the central
binary. This is inside the chaotic zone around the stellar binary
because the zone radius for a twin corresponds to $f \approx 3.7$
(see \citealt{HW99} and Figure~2 in \citealt{S15}). If one takes
$Q_\mathrm{p}' = 100$ instead of 500, then the habitability range
in $f$ would be 2.4--4.0; and if, in addition, one takes
$R_\mathrm{p}=3 R_\mathrm{Earth}$, then the range in $f$ would be
2.8--4.7. In both cases, the zone of suitable tidal heating
overlaps with the zone of orbital stability. Taking into account
the highly approximate character of the model, one may conclude
that the forced eccentricity of CBP can indeed be an important
factor in maintaining active tectonics on such planets, and there
can even be no need for radiogenic heating to maintain them
tectonically active, contrary to the case of Earth.

Finally, note that the equilibrium ratio of planetary rotation and
orbital frequencies is equal to $(1 + 19 e^2 /2)$ and $(1 + 6
e^2)$ in the constant phase-lag and constant time-lag tidal
models, respectively; see, e.g., \cite{B15}. Therefore, the forced
orbital eccentricity also solves the problem of life-unfriendly
synchronous tidal locking. Such a locking implies that the central
illumination source permanently faces one and the same side of a
planet; planets in the HZ of single M-dwarfs are especially prone
to this effect --- see \cite{SBJ17}. Due to the forced
eccentricity, for CBP this effect is impossible.

In conclusion, the tidal heating may provide a natural automatic
constant internal heating for CBP-MS. However, its effectiveness
deserves a further study. Conversely, the heating of the Earth by
Lunar tides is accidental, not generic, having resulted from the
Moon-forming impact.

\section{Protection from stellar wind}
\label{sec_ProtSW}

\cite{M13,M15,M15b} came to the conclusion that CBP might be
exceptionally good, in an important astrophysical aspect, for
abiogenesis: in the main-sequence stellar binaries with orbital
periods longer than $\sim$10~d and less than $\sim$50~d, the
stellar rotation and binary revolution synchronize with each
other, and the tides on the binaries radically suppress the
magnetic dynamo mechanism in the host stars, thus reducing
chromospheric activity and life-hostile extreme UV and stellar
wind. This mechanism reduces ``stellar aggression'' and favors
retaining moist atmospheres for the planets in habitable zones.
The planetary atmospheres are protected from mass loss and water
loss, and strong magnetospheres are not needed, contrary to the
case of planets of single stars.

The given ``Binary Habitability Mechanism'' (BHM), as coined by
\cite{M15}, is effective for the stellar binaries with periods
from $\sim$10~d up to $\sim$50~d \citep{M15}. It is easy to
calculate that the upper bound corresponds to the orbital period
$\sim$250--350~d for a planet in resonance cells 5/1--6/1 or
6/1--7/1, at the border of the chaotic zone around the central
binary. Such a period is similar to Earth's orbital period, and
thus the planet turns out to be close to the habitable zone of
Solar-like host stars. Conversely, taking the lower bound
$\sim$10~d for the period of a BHM-effective stellar binary, one
finds that it corresponds to the period $\sim$50--70~d for a
planet at the chaos border. Therefore, the planet appears to be
tentatively inside the habitable zone of M-class host stars, as,
according to \cite{Q14}, {\it Kepler}-186f is.

While the protection of CBP-MS from stellar wind is provided, due
to the BHM, almost automatically, Earth's permanent magnetospheric
protection might have been granted by an accidental event in its
early history, namely, by the impact of a planetary embryo: this
impact enriched the planet with iron, and a substantial metallic
core was formed (see \citealt{T93} and references therein). On the
other hand, the same impact is thought to have been responsible
for the formation of the Moon.

However, note that some doubts exist on whether the relevant
properties of binary stars are indeed favorable. In \cite{JZP15},
it is argued that the protective magnetospheres of CBP may
decrease in size because the planets may suffer strong shock waves
several times per orbital revolution.

\section{Delivery of water}
\label{sec_Dwater}

The problem of appearance of water on the Earth is most
disputable, and, consequently, any inferences on this habitability
condition can only be rather speculative. In one of the scenarios,
water is thought to have been delivered to the Earth only after
its formation, as the volatiles are lost during the
formation~\citep{MCL00}. According to the Nice model
\citep{GLT05}, the delivery of water to the after-molten Earth was
partially due to a fortuitous event: the passage of a migrating
Jupiter and Saturn through the 2/1 mean motion resonance. This
passage provoked disturbances in the planetesimal debris disks,
and swarms of icy minor bodies entered the inner realms of the
Solar system, bearing water to the already-cooled Earth.

For CBP-MS, an analogous process may emerge due to the passage of
a migrating planet through the 6/1 or 7/1 or a higher-order mean
motion resonance with the central binary, before the migration
stalls the planet at an appropriate resonance cell on the border
of the central chaotic zone. This passage provokes disturbances in
the outer parts of debris disks, and icy bodies may enter the
inner realms of the circumbinary system, bearing water to
already-cold planets. The process is automatic, in contrast to
that in the Solar system.

However, in the scenario believed at present to be the most
plausible, most of the water is delivered to the Earth by a few
planetary embryos from the outer asteroid belt at the final stage
of Earth's formation, and only a small fraction (less than 10\%)
is delivered by later bombardments \citep{MCL00}. This scenario is
verified by the observed isotope ratios (the D/H ratio first of
all) in Earth's seawater and in meteoritic and cometary material.
Data retrieved from studies of the Jupiter family comet
67P/Churyumov--Gerasimenko by the {\it Rosetta} spacecraft shows
that its D/H ratio is three times the terrestrial value
\citep{ABB15}.

The percentage of water delivered to the Earth by later
bombardments does not exceed 10\% \citep{MCL00}. That is why the
more or less automatic mechanism of water delivery from outer
reservoirs (stellar Oort clouds) by the Galactic tide cannot alone
provide the water amounts necessary for habitability. Therefore,
mechanisms acting during planetary formation and/or migration in
the system are needed. Clearly, statistical preferences for the
delivery of water to CBP (in comparison to planets of single
stars) require further study.

\section{Ecliptic variations}
\label{sec_Eclipses}

Neap/spring tides arise due to the beating between the Solar and
Lunar tides. This beating is substantial because the amplitudes of
the Solar and Lunar tides are of the same order of magnitude.
Thus, a new, longer timescale is introduced in the tidal
variations. As outlined by \cite{B01}, for the tidal beating to be
present on a planet, the angular sizes of the perturbing bodies,
as seen from a planet, should be roughly equal.

Consequently, the existence of the tidal beating on CBP-MS is
generic, and, what is more, in coplanar systems it is accompanied
by prominent stellar eclipses --- periodic rapid changes of the
stellar flux received by the planetary surface.

Therefore, there exists an {\it ecliptic analogy} between CBP-MS
and the Earth: in both cases, the stellar eclipses are common.
However, there exist two substantial differences. (1)~In the
former case, the ecliptic shadows are typically global: the
planetary star-side surface can be shadowed by a stellar companion
totally, whereas the Moon's shadow traces the Solar side of the
Earth only partially. This means that the ecliptic factor in
abiogenesis, if any, may play a greater role in CBP-MS than in
Earth. (2)~On CBP-MS, the flux never reduces to zero, as the
eclipsing body also radiates. The reduction magnitude depends on
spectral classes of the binary components. The relative minimum is
achieved when the eclipse is complete but the disk of the
transiting star, as seen from the planet, is not much greater than
the disk of the eclipsed star, so that the luminosity of the
transiting star is minimal. For a planet in resonance cell
5/1--6/1, the eclipse of a G2V (``Sun''-like) component of the
binary by its K5V counterpart (``61~Cyg~A''-like; note that
61~Cyg~A has radius 0.665, luminosity 0.153, and mass 0.67 of the
Solar values, \citealt{KMP08}) would imply a fourfold reduction of
the total flux at the planetary surface. Besides, the spectrum of
the received radiation varies during the eclipse quite radically.

As the eclipses provide sharp and fast changes in the surface
temperature, these frequent and periodic phenomena may modify the
photo-tidal driving mechanism of PCR-type considered in
Section~\ref{sec_Tides}.

Photoperiodism and photosynthesis conditions on CBP were
numerically explored in \cite{FMC15}; a rich variety of ``forcing
timescales for photosynthesis'' was recovered, which is appealing
for further studies of possible biospheres ``rich in rhythms and
cycles.'' Frequent and periodic eclipses induce an even greater
diversity. CBP-MS, whose orbits are coplanar with orbits of their
parent stellar binaries, are more able to provide such a diversity
in this respect than Earth because the eclipses are frequent and
generic on them; in contrast, on the Earth they are rare and
non-generic, being conditioned by a chance orientation and size of
the Lunar orbit.

Quite a perfect alignment of the orbital planes of a CBP and its
parent stellar binary is needed for this ecliptic mechanism to
work. However, note that the alignment in the most {\it Kepler}
circumbinary systems is indeed perfect: the mutual inclination of
stellar and planetary orbits typically does not exceed
0.5$^\circ$~\citep{W12,OWC12a,OWC12b}.

\section{Discussion}
\label{sec_Discussion}

In this section, open and controversial issues in the proposed
scenario are discussed.

\subsection{Formation issues}
\label{subsec_Formation}

The observed CBP-MS all have masses in the range from 0.1 to 0.5
Jovian masses~\citep{D11,OWC12a,OWC12b,W12}. Therefore, they are
Neptune-like or Jupiter-like. They cannot be Earth-like or cannot
even belong to the class of super-Earths. However, it is clear
that the observed large-mass preference does not refute, due to an
obvious observational bias, the potential existence of smaller
rocky planets in the same systems because the transit signal from
a super-Earth or an Earth-like planet is $\sim$100--1000 times
smaller than that from a Jupiter-like planet. As discussed in
Section~\ref{sec_Insolation}, the smaller planets can reside in
the orbits outer to the observed ``leading'' CBP, but basically
not inner to them because the inner orbits are mostly unstable.

Other possibilities include the following. (1)~Large satellites of
the observed CBP-MS may be habitable \citep{QMC12}. However, this
opportunity requires a strong revision of the whole scenario, in
application to satellites' physics and dynamics. (2)~The observed
CBP-MS can be non-habitable, but the co-orbital material that they
shepherd can be habitable. The co-orbital material may include any
rocky objects of suitable masses; a particular class of such
objects was considered as ``Trojan exomoons'' in \cite{QMC12}. The
dynamical stability of the material co-orbital with the planet
sharply increases if the host star is binary~\citep{DS16}.

Since super-Earth or Earth-like class CBP-MS have not yet been
discovered, let us consider whether the modern theories of planet
formation favor their actual existence.

Though a number of problems still remain enigmatic, the
terrestrial planet formation in single-star systems is a
profoundly explored subject
\citep{R06,R08,RQL04,RBK06,RQL06,RSM07,ROM09,DAA14}. Conversely,
the terrestrial planet formation in circumbinary systems is quite
a novel subject, given that actual CBP were discovered only
recently.

Terrestrial planets and cores of giant planets have a common
origin: they form in protoplanetary disks containing dust and gas
\citep{S69,L93}. The formation has three distinct stages
\citep{S69,L93,R08}: (1)~dust grains in the gas--dust
protoplanetary disk coalesce into km-sized planetesimals; (2) the
planetesimals form thousand-km planetary embryos via pair-wise
accretion; and (3) the embryos form full-sized planets via merging
impacts.

Stage~(1) is a complex physico-chemical process, whose details are
not all established yet, but there is no known reason to believe
that it is more difficult in circumbinary disks than in disks of
single stars. Stage~(3) proceeds effectively in circumbinary
disks, producing varieties of multiplanet orbital configurations
\citep{QL06,QL10}. \cite{QL06} performed numerical simulations of
this stage for a circumbinary disk surrounding short-period binary
stars; they found that building terrestrial planets, starting from
Moon-sized planetesimals, in circumbinary environments is not much
more difficult than in single-star systems, and a diversity of
circumbinary terrestrial planets can be generated, given that
large building blocks are present.

It is stage~(2) that looks most controversial. Before this stage,
km-sized planetesimals emerge, large enough for the gravity forces
to become dominant in their dynamics in comparison with the
aerodynamical drag in the gas disk. For CBP formation, the main
problem is just with stage~(2), in which the pair-wise accretion
of the km-sized planetesimals takes place. \cite{MN04} showed
analytically that an extended (with respect to the central cavity)
circumbinary zone is hostile for pair-wise planetesimal accretion.
The barycentric radius of this unfriendly zone is typically by an
order of magnitude greater than the radius of the central chaotic
zone. The accretion is precluded because the stellar binary drives
the eccentricity and differential (radius-dependent) precession of
the planetesimal orbits; this implies destructive collisions of
planetesimals. Only quite far from the barycenter, the collisional
velocities become small and start to allow accretion. In a
gas-containing disk, the eccentric orbits forced by the central
binary would be aligned apsidally due to the gas drag, but the
drag depends on planetesimal size, and this introduces
differential orbital phasing, increasing the collisional
destruction of bodies of different sizes \citep{TMS06}.

Within the accretion framework, the following scenario for the
formation of CBP is favored \citep{PN07,M12,PLT12}: the planetary
core forms in the outer accretion-friendly zone in the
protoplanetary disk; then it migrates inward and stalls at the
border of the central cavity. \cite{PLT12} performed simulations
of planetesimal evolution in circumbinary disks, including the
formation of km-sized planetesimals; it was found that {\it
Kepler}-16b, 34b, and 35b are unlikely to have formed in situ.

As \cite{BK15} note, in the outer regions of circumbinary disks,
planets form ``in much the same way'' as their analogues in
single-star systems. The circumbinary environment may even be
friendly to planet formation. However, migration through viscous
interactions with a gas-containing disk seems to be a necessary
factor in forming final architectures. The location of the orbits
of the observed giant planets in {\it Kepler} systems are
reproduced at certain parameters of viscous drag and disk profile
models \citep{PN13,KH14,KH15}.

\cite{N03} was first to study the migration of CBP in disks; his
hydrodynamical simulations showed that a Jupiter-like giant can be
captured in the 4/1 resonance with the central binary. \cite{PN07}
showed that lower-mass CBP (with masses from 5 to 20 in Earth
units) may stall at the border of the disk cavity and predicted
that Neptune-like CBP can be discovered there. This was confirmed
by discoveries of the {\it Kepler} CBP. An important role of
orbital resonances was considered in \cite{PN08a,PN08b} in
hydrodynamical simulations of planetary migration in circumbinary
gaseous disks. Such resonances as the 4/1 and 5/1 with the central
binary may serve as traps for migrating planets and generally
influence the formation process and orbital evolution of giant
Saturn-mass planets embedded in the disk. These results, favoring
inward migration, were confirmed in \cite{KH14}, where a detailed
balance of viscous heating and radiative cooling of the disk was
taken into account, and in \cite{LLB16}, where planetesimal growth
was considered taking into account perturbations due to the
precession of the eccentric disk.

\cite{PN07} were first to explore the outcomes of type I migration
of low-mass planets in gaseous circumbinary disks. It was found
that the migration can be halted at the edge of the central
cavity. For a gaseous disk, this halting is explained by
counterbalancing the negative Lindblad torque by the positive
corotation torque; this ``planet trap'' mechanism (considered
below in more detail) operates when there is a sharp positive
gradient of the disk surface density \citep{MMC06}. \cite{PN08a}
performed hydrodynamical simulations of multiple-planet systems
embedded in circumbinary disks. In a system consisting of two
planets, when the largest one has settled at the cavity border and
the smaller one continues to migrate inward from outwards, the
outcome usually represents an equilibrium configuration with the
planets locked in orbital resonances. (If the smaller planet is
first to settle at the cavity border, the usual outcome consists
of scattering of the planets.) In a simulation of a five-planet
system, \cite{PN08a} revealed a stable final outcome representing
a totally resonant three-planet system. This is in accordance with
the possible existence of smaller planets outer to the observed
``leading'' giants, as proposed in Section~\ref{sec_Insolation}.

All mentioned results of massive hydrodynamical simulations are in
agreement with analytical predictions of the so-called ``planet
trap'' concept, developed in \cite{MMC06} to overcome the problem
with type I planet migration, which can be too fast relative to
the rate of planet build-up, thus eliminating the planet totally.
As already mentioned above, when there exists a sharp positive
gradient of the disk surface density at some radial distance from
its center, the negative Lindblad torque can be counterbalanced by
the positive corotation torque, and the planet-forming material
can be trapped at this location \citep{MMC06,M14}. \cite{MMC06}
identify possible locations of the traps: apart from the edge of
the central cavity, these locations may comprise the outer borders
of resonant ring-like zones cleared from the gas by giant
protoplanets (the single-star case was considered, but the
circumbinary case is analogous and is even more representative).
The smaller planets outer to the observed ``leading'' giant CBP
can, in principle, be trapped at these local borders.

In the framework of the planetary trap concept, \cite{M14}
suggested that large planetesimals form at the cavity edge, where
the pressure is maximum and the surface density of solid material
is enhanced; this may give rise to CBP formation in situ.
Circumbinary planet formation in systems of close binaries may be
even easier than planet formation around single stars
\citep{MAA13}. Indeed, it is usually assumed that the circumbinary
disk is fully ionized and turbulent, but in fact, it is typically
layered, consisting of a radially extended ``dead zone'' (a
non-turbulent midplane layer) and two surface layers (strongly
turbulent). The dead zone is a place favorable for planet
formation where solids may settle to the midplane. Therefore,
according to \cite{MAA13}, in poorly ionized disks, CBP may form
in an easier way than planets in disks of single stars.

The radial locations of the planet traps in single-star disks are
ill-constrained \citep{M08}, and this fact, as outlined in
\cite{PPZ13}, is the principal problem with the concept of planet
traps in single-star disks. The circumbinary case is much more
clear-cut in this respect because circumbinary protoplanetary
disks possess a definite scale defined by the size of the binary.
Therefore, the radial location of the sharply positive radial
gradient of the surface density is perfectly defined. As opposed
to single-star disks, circumbinary disks do not need excessive
fine-tuning of parameters to provide suitable radial locations of
forming planets.

For single stars, \cite{MLO12} emphasize a great diversity in the
evolved orbital architectures of terrestrial planets, given that
the accretion process for the forming terrestrial planets depends
sensitively on the orbital configuration and evolution of giant
planets, which form first in the protoplanetary disk. Can an even
greater diversity be expected in planetary systems of binary
stars? For the S-type (circumcomponent) planetary systems in wide
binaries, this can be indeed so, but for the P-type (circumbinary)
systems, the inference is not obvious at all. Formally comparing a
circumbinary planetary system architecture with that of the Solar
system, one sees that the former may be considered as a
``truncated version'' of the latter: the inner rocky part of the
system simply does not exist because it is in the circumbinary
chaotic zone. In most known CBP-MS, this zone is closely orbited
by a giant planet, a Jupiter analogue. If this ``truncated
system'' picture is valid, then the orbital architecture of the
zone outer to the leading giant can be rather simple, resembling
that of the Kuiper belt, with icy/rocky objects tending to reside
in orbital resonances with the shepherding giant. (The latter's
role in the Solar system is played by Neptune.) This architecture
is in accordance with the possible existence of terrestrial CBP
outer to the leading giant CBP, as proposed in
Section~\ref{sec_Insolation}.

An observational evidence for the existence of planets outer to
those trapped at the edge of the central cavity is provided by the
discoveries of such planets as {\it Kepler}-47c, {\it
Kepler}-1647b, and OGLE-2007-BLG-349L(AB)c
\citep{OWC12a,KMH13,HHK15,K16a,BRU16}. The discoveries of such
objects demonstrate that not all CBP halt migration at the cavity
edge, but some settle at much greater distances from the host
binary. In the case of the {\it Kepler}-47 multiplanet system,
there is a lot of space between {\it Kepler}-47b and {\it
Kepler}-47c to harbor possible dynamically stable low-mass planets
\citep{HHK15}, i.e., the observed two-planet system is in no way
closely-packed dynamically. The systems {\it Kepler}-1647 and
OGLE-2007-BLG-349L(AB) may as well contain low-mass planets in
orbits inner to the observed giant planets because the orbits of
the former ones are large in size in comparison with the central
chaotic zone.

In conclusion, from the viewpoint of planet formation scenarios, a
major preference for the occurrence of CBP in locations suitable
for habitability seems to exist indeed. In brief, in circumbinary
systems, the major planet trap forms at the strictly defined (in
radial location) border of the central cavity in the
protoplanetary disk. Due to the observational preference for
particular values of the orbital periods of twin Solar-like
binaries (this preference also has a theoretical explanation, as
described in Section~\ref{sec_Insolation}), any planets moving
just outer to the leading giant CBP are preferentially placed
inside the insolation HZ. In other words, there exist a natural
radial scale and pattern for the overall planetary architecture in
circumbinary systems, whereas there are no such clear-cut scale
and pattern in single-star systems, at least as known at present.

\subsection{Non-ubiquity of double planets}

Terrestrial CBP, as discussed above in
Subsection~\ref{subsec_Formation}, seem to be generic, whereas, on
the other hand, double planets seem to be non-generic: the
probability for an Earth-like planet to acquire a large Moon-like
satellite, from the cosmogonical viewpoint, is extremely low
\citep{L97}.

The theme, however, remains disputable, as recent massive
numerical experiments taking into account fragmentation and
bouncing collisions show. In \cite{QBB16}, model statistics of
giant impacts during the late stages of terrestrial planet
formation around a Solar-like star was obtained, according to
which the impacts forming Moon-sized satellites are rather typical
in the protoplanetary systems of Solar-like stars. Therefore,
double terrestrial planets are not necessarily very rare. Clearly,
the problem of ubiquity of double terrestrial planets in habitable
zones of single stars requires further quantitative study,
including estimates of the abundance of such objects.

\subsection{Exomoons}

Habitable exomoons may exist under appropriate conditions
\citep{K10}. Exomoon habitability is constrained by orbital
stability, illumination, and tidal heating \citep{H12,HB13}.
Magnetic shielding of exomoons can be provided by the
magnetospheres of the host planets \citep{HZ13}. \cite{HiK13}
considered habitability properties of exomoons residing either far
from their host planet, close to the Hill stability border, or
conversely, close to the planet, near the tidal locking threshold.
Exomoons can be quite different from the moons present in the
Solar system; in particular, they can be much larger in size. As
\cite{HP15} found in numerical simulations of the cosmogony of
satellites of super-Jupiters, Earth-size moons may form at
appropriate (permitting the habitability) orbital distances from
the host planet.

Like any other planets, CBP-MS may have moons inside their Hill
spheres. Since all known CBP-MS are giants, these moons can be
large and can, in principle, be habitable. Exomoon habitability
conditions were considered in \cite{HP15}. These conditions are
modified by the binarity effect of the host star. \cite{QMC12}
considered hypothetical moons of {\it Kepler}-16b as possible
habitable niches, as {\it Kepler}-16b itself is a Saturnian-type
giant planet and may hardly be habitable. \cite{F14} suggested
that while {\it Kepler}-47c (a planet in the HZ of {\it
Kepler}-47) is a giant one (not terrestrial), it may have a
habitable terrestrial satellite, with the climate subject to
strong variations.

Generally, the expected strong day/night, seasonal, and climatic
variations on possibly habitable terrestrial exomoons on various
timescales related to the orbital motions of the host planet and
the moon itself can make the survival of their biospheres
questionable. For exomoons of CBP, where additional variations
related to the orbital motion of the host stellar binary are
present, the picture becomes even more complicated. As \cite{F14}
notes, if a giant CBP in the stellar HZ, such as {\it Kepler}-47c,
has an Earth-like moon, the latter's biosphere can be sustained
only if it is able to endure strong climatic variations.

\subsection{Spin--orbit coupling}

An analysis by \cite{C15} demonstrated that stable spin--orbit
coupling may well exist for circumbinary bodies in circular orbits
in the hierarchical three-body problem. However, as shown by
\cite{C15}, for gas giants, the coupling is unlikely due to their
expected almost perfect axial symmetry, similar to that of Jupiter
or Saturn; for terrestrial CBP, the coupling may be plausible
because the asymmetry is greater.

Spin--orbit resonances generally do not preclude habitability.
\cite{BMF14} studied photosynthetic conditions in a model of a CBP
in 3/2 spin--orbit resonance and concluded that life on such a
planet may well survive, though, contrary to the Earth, it would
have geographically longitudinal, not latitudinal, areas of
habitat. Capture of a CBP in low-order spin--orbit resonances,
such as 3/2 or 1/1 (synchronous) ones, is not expected to affect
the photo-tidal synchronization; such resonance would only mean
elimination of fast (day/night) insolation and tidal harmonics.
Thus, the photo-tidal synchronization would operate in its sheer
form. The interplay of fast and slow photo-tidal harmonics, when
CBP rotates rapidly, deserves a separate study.

\subsection{Photosynthetic efficiency}

Not only the total flux of stellar radiation at the surface of a
planet matters, but the radiation spectrum as well. If the
companion of a yellow dwarf in a binary is a red one, this can
favor a variety of primordial chemical reactions on planetary
surface, since the spectrum of the light falling on the planet is
significantly widened. As mentioned by \cite{M13}, an excessive
photosynthetic radiation flux provides an additional excellent
condition for habitability of CBP.

It is well known that the absorption spectrum of chlorophyll is
double-peaked, and the peaks do not coincide with the maximum of
the Solar spectrum; see, e.g., \cite{SBJ17} and references
therein. Since there exist evolutionary (rather complicated)
scenarios explaining this spectral mismatch (see a review in
\citealt{RW04}), it would be premature to interpret the
double-peaked absorption spectrum of chlorophyll as a
``fingerprint'' of a double star. However, note that if the
components of a binary are even only slightly different in
spectral class, the widening of the composite spectrum can
increase the photosynthetic quality of the radiation received at
the surface of a CBP with Earth-like photosynthesis.

The photosynthetic conditions at some known CBP-MS were modeled in
detail in \cite{FMC15}; they explored the variations of the
photosynthetic spectral quality of radiation with time and surface
position. The photosynthetic conditions at CBP provide a much
wider field for evolutionary diversification than on Earth.
Combinations of G and M stars provide opportunities for both
familiar and exotic forms of photosynthetic life, such as infrared
photo-synthesizers \citep{ORC12}.

\subsection{Habitability of CBP of M-dwarf binaries}

As noted already in Section~\ref{sec_Insolation}, M-dwarfs can be
especially suitable as host stars of habitable CBP. Habitability
properties of planets of M-dwarfs were extensively studied in the
last years; see reviews by \cite{TBM07} and \cite{SBJ17}. While
M-dwarfs are numerous and have long lifetimes in comparison with
stars of other spectral classes, two major disadvantages for the
potential habitability of their planetary systems
exist~\citep{CG16}, namely, (1)~small sizes of stellar habitable
zones and (2)~frequent flares, including superflares. Note,
however, that for circumbinary systems the situation can be
different. The circumbinary HZ is more extended, as a twin binary
produces twice as many photons as any of its components, and of
the same spectrum (see Section~\ref{sec_Insolation}); what is
more, the magnetic activity can be suppressed via BHM, discussed
in Section~\ref{sec_ProtSW}.

Planets of M-dwarfs have on average smaller sizes than planets of
Solar-like stars~\citep{MCS17}. This observational fact may
provide another answer to the problem of formation of planets of
terrestrial type, again pointing out that abiogenesis can be
concentrated on CBP of double M-dwarfs.

M-dwarfs comprise more than 70\% of the Galactic stellar
population, and more than 50\% of the stellar population are in
binaries \citep{DM91,BHC10}. The mass function of M-dwarfs peaks
at classes M3V--M4V \citep[Figure~23]{BHC10} --- just where
resonance cells 5/1--6/1 and 6/1--7/1, typically occupied by
planets, overlap with the circumbinary habitable zone (see
Section~\ref{sec_Insolation}). This means that if abiogenesis is
indeed concentrated on CBP of double M-dwarfs, then the production
of replicating biopolymers in the Galaxy can be a massive
phenomenon.

\subsection{Abiogenesis apart from TCR}
\label{subsec_aaft}

In this article, we have focused almost entirely on Lathe's tidal
mechanism for abiogenesis because it has directly verifiable
astronomical aspects. Let us briefly discuss the astronomical
aspects of other modern approaches to the problem of life origin.

The photo-tidal periodic driving could also be important in the
``everything-at-once'' abiogenesis scenario, where metabolism,
replication, and compartmentalization all arise simultaneously.
The scenario is described in \cite{PGS09}, \cite{Su15}, and
\cite{PPR15}. According to it, the first life forms are thought to
arise in pools generated by hot metal-enriched water streams in
tectonically active environments. Analogously to Lathe's
mechanism, an intermittent cooling/heating is needed to provide
association/replication cycling of the proto-replicating material,
pumping its concentration. Heating induces the strands of the
replicating material to come apart.

An important point is that the compartmentalization of
proto-replicating material in protocells may provide extended
timescales for keeping the produced biomaterial from destruction;
therefore, the problem of a necessarily short period of the
association/replication cycling in Lathe's mechanism (discussed in
detail in \citealt{L05}) can be avoided. However, the period of
the proto-replication cycle should be somehow adjusted to the
astronomical photo-tidal cycling period.

Therefore, Sutherland et al.'s geochemical scenario seems to be
subject to the same general photo-tidal conditions as those in
Lathe's mechanism, and therefore, it may operate on CBP-MS in the
same way as it hypothetically operated on Earth.

Of course, apart from the photo-tidal ones, other conditions
should be satisfied, such as active tectonics, UV radiation, and
in particular, the presence of particular metals (copper, iron,
sulphur, magnesium; \citealt{PGS09,PPR15}).

Concerning active tectonics, on CBP-MS it can be provided
automatically, as we have seen in Section~\ref{sec_Active_tect},
at least for a subclass of CBP-MS. What is more, on CBP-MS, the
available UV radiation is subject to a periodic variation with
multiple harmonics, some of which might be suitable for
abiogenesis.

In conclusion, Lathe's TCR mechanism \citep{L04,L05} for
abiogenesis on the Earth is favored as generic for CBP-MS due to
the photo-tidal synchronization that is inherent to them. For the
same reason, CBP-MS can also be suitable for Sutherland et al.'s
``multi-stream'' abiogenesis.

Martin and Russell's abiogenesis scenario (see \citealt{K12} and
references therein), in which life is born in deep-sea
hydrothermal vents, is indifferent to the presence of the
photo-tidal cycling, at least in its ``photo'' constituent. Since
CBP-MS do not provide any theoretical preference here, one may
expect that any verification of this scenario also checks the
importance of the photo-tidal cycling for abiogenesis.

\subsection{Superhabitability}

Heller \& Armstrong (2014, p.~50) outline ``the possible existence
of worlds that offer more benign environments to life than Earth
does,'' and define such worlds as superhabitable.

According to \cite{HA14}, terrestrial planets with masses from
$\sim$2 to $\sim$3 Earth masses are potentially superhabitable due
to a number of circumstances, namely due to (1)~the prolonged
tectonic activity, and as a consequence, the prolonged existence
of the carbon--silicate cycle, (2)~the enhanced magnetic shielding
owing to a greater mass of the iron core, (3)~the greater surface
area resulting in greater biomass production, (4)~a smoother
surface providing shallower seas, and (5)~the ability to maintain
thicker atmospheres.

On the other hand, K-dwarfs, being only a little bit less massive
and luminous than the Sun, have much longer lifetimes, thus
favoring long-term habitability of their planets.

Combining these considerations with our inferences for CBP, one
may expect that the CBP belonging to the super-Earth class and/or
orbiting K-dwarf binaries can superhabitable, in the definition of
\cite{HA14}. However, this conclusion should be put in
correspondence with our inferences made above for Solar-like stars
and M-dwarfs as host stars; we leave it for future work.

\subsection{Intragalactic transport}
\label{subsec_IGt}

Finally, it is interesting to note that circumbinary systems may
give birth to rogue planets transporting the masses of produced
biopolymers or any other biogenic chemicals elsewhere. On
cosmogonical timescales, the process of escape of planets from
host stellar binaries seems natural because the mass ratio and
size of the binary inevitably change with time. This evolution is
due, in particular, to the stellar mass loss and mutual tides
\citep{HTP02}. Therefore, a CBP-MS, initially automatically
stalled at the edge of the central chaotic zone, may enter it and,
consequently, escape. This would entail the destabilization and
reconfiguration of the overall planetary system, including the
possible scattering and ejection of its other planets. A
production of rogue planets by stellar binaries was considered by
\cite{VT12} in similar dynamical scenarios.

The final escape of planets seems to be natural in the case of
circumbinary systems, but not for our Solar system, which is known
to be extremely stable. Indeed, only Mercury may become rogue, as
revealed by \cite{L94}, on a timescale of several billion years.

In fact, single-star and binary-star populations can both loose
their planets by a number of dynamical and physical mechanisms
(stellar fly-bys, planet--planet scattering, supernova explosions,
etc.; see \citealt{VT12,VEW14,K16b}). No comparative planetary
escape statistics is available up to now, due to the complex and
multifaceted nature of the phenomenon. However, CBP (at least the
CBP orbiting at the edge of the circumbinary chaotic zone) seem to
be much more prone to any destabilizing factors. \cite{SF16},
based on massive numerical simulations, find that the planetary
escape process from circumbinary systems may efficiently fill the
Galaxy with rogue planets. Most CBP are ejected rather than are
destroyed via planet--planet or planet--star collisions
\citep{SF16,SKS16}.

The numerical study by \cite{K16b} of the long-term dynamical
evolution of the {\it Kepler} circumbinary systems, taking into
account the stellar evolution of the host binaries, reveals that
the expansion/contraction of CBP orbits can be adiabatic, coherent
with the evolution of the host binary. This reduces the chances
for CBP ejection. However, at least two {\it Kepler} binaries
(namely {\it Kepler}-34 and {\it Kepler}-1647) are doomed to free
their planets because in the course of stellar evolution, they
experience a double-degenerate supernova explosion. Tantalizingly,
these two binaries belong to the class of twin binaries, i.e., the
stellar companions are quite similar in mass in each of the
systems.

In conclusion, the intragalactic migration and dissemination of
the planetary chemicals can have two sharply different modes,
corresponding to the multiplicity of host stars. CBP-MS can be
mostly ejected in the course of the stellar evolution of their
host binaries and become rogue and then migrate freely, whereas
planets of single stars are generally orbitally superstable (as
the Earth is) and thus are doomed to migrate radially slowly
together with their host stars (on stellar migration, see
\citealt{SB02,MF10,S11,GSK16}).

\subsection{Methods}

Some basic methods used in this article concern the analysis of
resonances in Hamiltonian dynamics. Namely, the extents of the
chaotic zones around gravitating binaries are estimated (in
Section~\ref{sec_Insolation}) using methods based on Chirikov's
resonance overlap criterion~\citep{C79,S15}. Theoretical resonant
and chaotic Hamiltonian dynamics form the basis for the obtained
inferences on the climate stability conditions
(Section~\ref{sec_Climates}), delivery of water
(Section~\ref{sec_Dwater}), and the final fate of CBP
(Section~\ref{sec_Discussion}). In Section~\ref{sec_Insolation},
the Lidov--Kozai effect again concerns resonant Hamiltonian
dynamics; on the resonant nature of this effect, see \cite{S17}.

\section{Conclusions}

Circumbinary planets are generic: indeed, a lot of circumbinary
planetary systems have been discovered up to now, and cosmogonical
simulations show that the formation of such stable systems is a
natural process, as discussed in
Subsection~\ref{subsec_Formation}. As we have seen, striking
analogies exist between the habitability conditions on CBP-MS and
on the Earth. In fact, in favoring the habitability conditions,
the Earth seems to mimic a typical CBP-MS. CBP-MS of particular
classes seem to be generic in providing such conditions (i.e., the
conditions arise automatically), whereas the Earth is not (i.e.,
the conditions arise accidentally). Therefore, the multiple
analogies revealed between CBP-MS and the Earth may indicate that
life on Earth is a low-chance outlier of a generic global chemical
process (massive production of replicating biopolymers)
concentrated on CBP-MS.

With respect to the insolation condition of habitability, at least
two habitability niches, where the needed insolation is provided
more or less automatically, can exist. These are the ``leading''
CBP of red-dwarf twin binaries and the ``outer'' (as defined in
Section~\ref{sec_Insolation}) CBP of Solar-like star twin
binaries.

In the considered scenario, Lathe's mechanism \citep{L04,L05} for
the tidal ``chain reaction'' abiogenesis on the Earth is favored
as generic for CBP-MS, due to the photo-tidal synchronization
inherent to them. For the same reason, CBP-MS can also be
favorable for Sutherland et al.'s ``multi-stream'' abiogenesis. On
the other hand, the scenario is indifferent to the deep-sea
hydrothermal-vent abiogenesis hypothesis.

In the Solar system, only one planet in the insolation HZ is known
to be biogenic. In the proposed scenario, this is accidental and
is mostly due to its duplicity, the Earth being an analogue to a
CBP-MS in several aspects. In circumbinary systems of MS twin
binaries, the HZ is broader than the HZ of single stars of the
same class, and, what is more, the planets need not be double to
be habitable; therefore, {\it all} super-Earths or Earth-like
planets in the insolation HZ can actually be habitable.
Consequently, in the proposed scenario, CBP-MS seem to be the main
carriers of biogenic chemicals in the Galaxy, whereas planets of
MS single stars produce biogenic chemicals only accidentally.

\section*{Acknowledgements}

The author is grateful to the referee for useful remarks. This
work was supported in part by the Russian Foundation for Basic
Research (project No.\ 17-02-00028) and the Programme of
Fundamental Research of the Russian Academy of Sciences
``Experimental and theoretical studies of objects of the Solar
system and planetary systems of stars.''

\end{document}